\documentclass[preprint]{vgtc}               

\ifpdf
  \pdfoutput=1\relax                   
  \usepackage{graphicx}                
  \DeclareGraphicsExtensions{.pdf,.png,.jpg,.jpeg} 
\else
  \usepackage{graphicx}                
  \DeclareGraphicsExtensions{.eps}     
\fi%

\graphicspath{{figures/}{pictures/}{images/}{./}} 

\usepackage{microtype}                 
\PassOptionsToPackage{warn}{textcomp}  
\usepackage{textcomp}                  
\usepackage{mathptmx}                  
\usepackage{times}                     
\usepackage{cite}                      
\usepackage{tabu}                      
\usepackage{booktabs}                  

\usepackage{amssymb}
\usepackage{pifont}
\newcommand{\cmark}{\ding{51}}%
\newcommand{\xmark}{\ding{55}}%

\onlineid{0}

\vgtccategory{Research}

\vgtcinsertpkg

\preprinttext{To appear in an IEEE VGTC sponsored conference.}

\title{Are LLMs ready for Visualization?}

\author{Pere-Pau Vázquez\thanks{e-mail: pere.pau.vazquez@upc.edu}}
\affiliation{\scriptsize ViRVIG Group - UPC, Barcelona}

\teaser{
  \centering
  \begin{tabular}{cc}
      \includegraphics[width=0.49\linewidth]{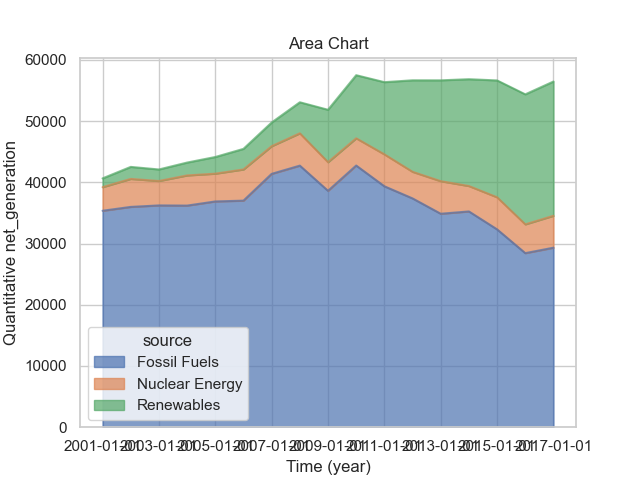} &  
      \includegraphics[width=0.49\linewidth]{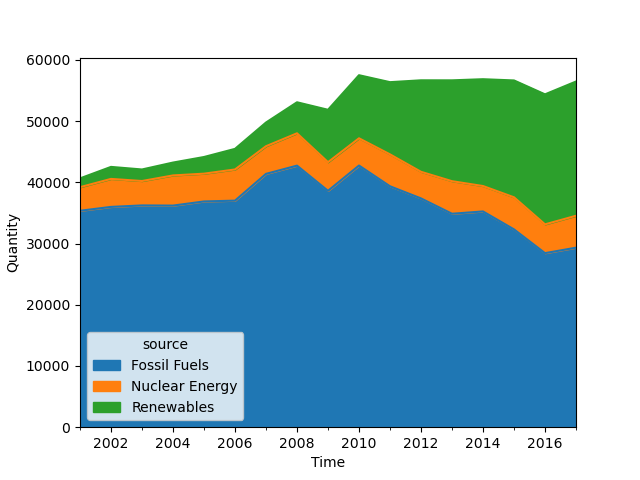} \\ 
       ChatGPT3 + matplotlib (default) & ChatGPT4 + matplotlib (default) \\ 
      \includegraphics[width=0.49\linewidth]{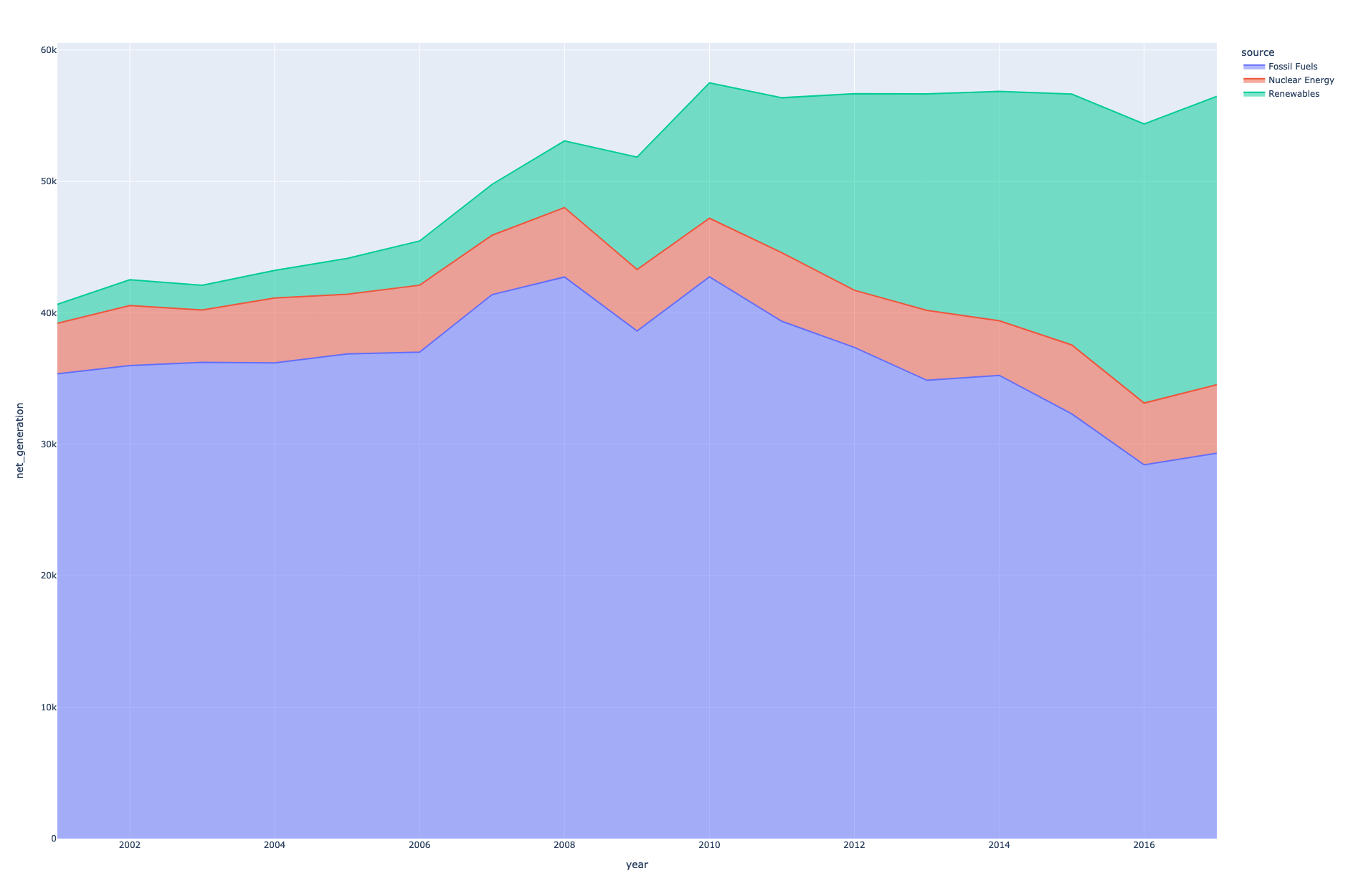} &  
      \includegraphics[width=0.49\linewidth]{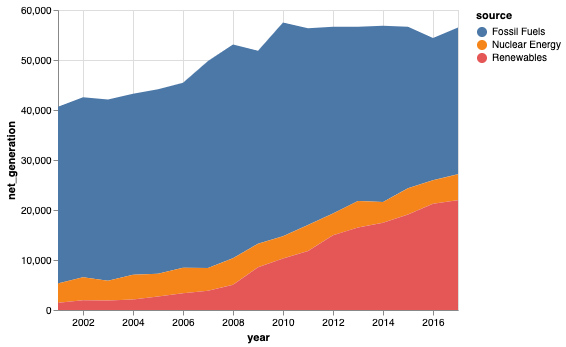} \\  
       ChatGPT4 + Plotly & ChatGPT4  + Altair \\ 
  \end{tabular}
  \caption{Generation of an area chart using ChatGPT3 and ChatGPT4. The top cases, the prompt did not ask for a concrete library. The bottom examples show the result when asking for a Plotly (left) or Altair (right). All the examples show the area chart, but ChatGPT3 generates labels that overlap. The solutions by ChatGPT4 seem all acceptable. The different configurations (legend placement and font sizes) are left to the libraries, which yield different looks. Also Altair (bottom right) sorts the areas in a different order.}
  \label{fig:teaser}
}

\abstract{Generative models have received a lot of attention in many areas of academia and the industry. Their capabilities span many areas, from the invention of images given a prompt to the generation of concrete code to solve a certain programming issue. These two paradigmatic cases fall within two distinct categories of requirements, ranging from "creativity" to "precision", as characterized by Bing Chat, which employs ChatGPT-4 as its backbone. Visualization practitioners and researchers have wondered to what end one of such systems could accomplish our work in a more efficient way. Several works in the literature have utilized them for the creation of visualizations. And some tools such as Lida, incorporate them as part of their pipeline. Nevertheless, to the authors' knowledge, no systematic approach for testing their capabilities has been published, which includes both extensive and in-depth evaluation. Our goal is to fill that gap with a systematic approach that analyzes three elements: whether Large Language Models are capable of correctly generating a large variety of charts, what libraries they can deal with effectively, and how far we can go to configure individual charts. To achieve this objective, we initially selected a diverse set of charts, which are commonly utilized in data visualization. We then developed a set of generic prompts that could be used to generate them, and analyzed the performance of different LLMs and libraries.
The results include both the set of prompts and the data sources, as well as an analysis of the performance with different configurations.
} 

\CCScatlist{
  \CCScatTwelve{Human-centered computing}{Visu\-al\-iza\-tion}{Visu\-al\-iza\-tion techniques}{};
  \CCScatTwelve{Human-centered computing}{Visu\-al\-iza\-tion}{Empirical studies in visualization}{}
}

\begin{document}


\firstsection{Introduction}

\maketitle

The generation of visualizations from natural language is not a new approach. For example, in 2001 there was already a first attempt by Cox et al.\cite{cox2001multi} to create visualizations from given commands. However, the progress, as in many other research fields, has rapidly evolved thanks to the public availability of Large Language Models. Previous research has focused on the creation of rules based on natural language input, and the inputs required a high degree of structure. This has undergone a transformation, particularly during the past decade, owing to the advancements in Natural Language Processing techniques \cite{Shen_2022}. The typical process for most systems consists of the description of the visualization creation commands in terms of data attributes and task \cite{narechania2020nl4dv}. This is especially useful when tackling data analysis tasks. Such systems are, therefore, primarily developed for visualization professionals. However, users may require a certain visualization technique for the available data. This could be of interest to not only visualization researchers and practitioners, but also to any type of user. It is an especially interesting situation given the increasing number of freely available open datasets. We envision the following scenario: a dataset is discovered on an open data platform, and we aim to generate a variety of visualizations of it, with the user having the ability to select the visualization technique employed. Currently, the obvious choice seems to be to use an LLM to generate such visualization, or the code needed to render it. However, it is unclear to what extent the outcome will be correct, both in terms of syntax and fulfilling the user requirements. This is the topic we intend to explore in this paper. 

We began the research by asking ourselves how good LLMs are at creating visualizations from a given dataset. This requires analyzing both the extent of their capabilities, such as the different techniques they can generate, and the depth of configurations that users can add to the desired charts. In this paper, we have conducted a systematic analysis of the capabilities of different LLMs by testing different, somewhat orthogonal features:
a) Variety of charts, b) Libraries that can be used, and c) Level of configuration of visual variables.

We contribute to the field in different ways: 
\begin{itemize}
    \item We have developed a prompt test set that can be used with current and future evolutions of LLMs. 
    \item We have analyzed how good are ChatGPT3 \cite{floridi2020gpt} and ChatGPT4 \cite{openai2023gpt4} in chart generation.
    \item We have analyzed the performance of ChatGPT4 with different libraries: matplotlib, Plotly, and Altair. 
    \item We have analyzed the capabilities of ChatGPT4 for configuring a limited set of visual variables in the most common chart techniques. 
\end{itemize}

The results are quite positive, with ChatGPT4 being able to create almost $80\%$ of the proposed charts. Furthermore, we have gained valuable insights into the capabilities of the LLMs to work with different libraries and data types.
All the prompts, datasets, and resulting code snippets analyzed are publicly available in GitHub\footnote{https://github.com/pere-pau/LLMsForDataVis}.




\section{Related Work}
\label{sec:Related Work}

The problem of using natural language to generate visualizations has been studied for more than 20 years now. The interested reader can refer to the recent surveys by Shen et al.\cite{Shen_2022} or Wu et al.\cite{Wu_2021}, that provide a comprehensive review of the most relevant developments throughout the last two decades.  
Cox et al. \cite{cox2001multi} created one of the first systems intended to generate visualizations from natural language. The field has undergone rapid advancements since the NLP researchers developed word embeddings, \cite{mikolov2013efficient}, which have proven to be a significant advancement in language comprehension. Although rule-based systems can be highly robust, they do not allow a significant amount of flexibility \cite{voigt2021challenges}. Other probabilistic grammar systems combine grammar rules and probability distributions to increase flexibility. Nevertheless, a high amount of expertise is necessary to generate such grammars \cite{narechania2020nl4dv}.

Many papers or systems focus on the use of natural language for data analysis through visualization. That is, visualization is a means to interpret the data. These systems must understand the user's intention  \cite{lee2012beyond} in order to craft the visualizations suitable to help them. As a result, significant efforts have been made to comprehend the analysis question. For example, Gao et al. focus on handling ambiguity in our language\cite{gao2015datatone}. Often, this issue cannot be automatically resolved and may require user intervention. In Eviza \cite{setlur2016eviza}, the authors address the problem in a task-based fashion, and deal with questions related to the data, such as "where is Texas" in a map. More recently, Narechania et al. adopted a different approach, named NL4DV \cite{narechania2020nl4dv}, a system that combines direct questions for visualization generation (e.g., "show me an area chart of how comedy movies evolved over the years") with follow-up questions (e.g., "As a boxplot now"). The system provides the user with the option to select the visualization, which is generated using Vega Lite \cite{satyanarayan2016vega}. However, the number of different available charts is also small. Other modern systems also provide a relatively small number of different visualization techniques \cite{Shen_2022}. Wang et al. \cite{Wang_2022} have created a NL framework to interpret the user's intentions in chart creation and edition. They demonstrate their architecture using a tool called VisTalk, which is an extended version of Excel that allows for verbal communication. The authors consider the system as a proof-of-concept, since it was created based on only a few samples. Furthermore, no indication was provided regarding the quantity and variety of charts that are accessible for manipulation.

In fact, the evolution of many of such NL2Vis systems has gone hand in hand with the evolution of Natural Language Processing technologies, as described by Shen et al. in their survey \cite{Shen_2022}. For example, Chen et al.\cite{chen2022type} use BERT \cite{devlin2018bert} to help interpret user inputs and create a domain-specific language for visualizations by fine-tuning a BERT model. Other systems have also taken advantage of LLMs to create visualizations. Two recent examples are Chat2Vis \cite{Maddigan_2023} and Lida \cite{Dibia_2023}. In both cases, the user requires an API key of some LLM (ChatGPT3.5, ChatGPT4, LlamaCode $\ldots$) to use the systems. Subsequently, these architectures utilize the LLM to interpret the input prompt and generate visualizations. This may pose a problem, since most people do not have a subscription to OpenAI LLMs and smaller LLMs do not produce high-quality results \cite{Dibia_2023}. Kim et al. conducted a very interesting experiment that analyzed the capabilities of ChatGPT of giving answers to visualization questions \cite{kim2023good}. Despite the results from the language model are not on par with humans, the authors expressed their optimism in the utility of LLM-guided advice. 
LLMs have also been used with a different perspective, such as chart explanations, or data exploration. Sultanum and Srinivasan have explored their use to support the authors of data-driven articles in providing explanations \cite{Sultanum_2023}. Furthermore, Lida also provides a first data summarizer stage in its pipeline \cite{Dibia_2023}, and a code interpreter, as well as visualization explanation, which are powered by LLMs.

In contrast to the systems that respond to user queries, other approaches have also focused on the fully automatic generation and recommendation of charts, based on data, such as \cite{Qin_2018,Ren_2021}. Our goal is to complement all of this research, since we are interested in the user directing the generation of visualizations. To this end, we evaluate how LLMs can be leveraged for data visualization generation, without API keys requirements, and all based on publicly available software.



\section{Overview of the process}
\label{sec:Overview of the process}

To evaluate the capabilities of LLMs, we had to determine the boundaries of our approach. Initially, and as previously mentioned, our objective is to enable the user to specify and delineate the desired outcome. This output could be either an image of a chart, or code. We opted for the second approach due to two factors, namely its utility and its evaluation. Firstly, generating code presents the possibility of modifying the output or using it as part of a more complicated framework. Second, it is easier for us to better understand how the prompt has been understood by the LLM if we further analyze the resulting code. Whether the system generated the appropriate code for reading or manipulating the data can only be answered if we fully inspect the output. We chose Python since it is a language that is expected to be highly represented in LLM training corpuses and has a wide range of visualization libraries. 

To create the prompt dataset and subsequent analysis, we performed the following steps:
\begin{itemize}
    \item Selection of visualization techniques.
    \item Creation or acquisition of datasets suitable for the techniques.
    \item Selection of LLMs to analyze.
    \item Design and fine-tuning of the prompts. 
    \item Exhaustive testing.
\end{itemize}

\subsection{Selection of visualization techniques}

The objective of the project was to evaluate the capabilities of LLMs in generating different visualization techniques. Therefore, a task-based approach (asking questions about the data) was inadequate. Instead, we needed to guide the LLMs to generate the techniques of interest. However, first, a set of techniques needed to be defined.

We approached this by using a progressive methodology. We first consulted Google and Google Images for charts. However, this results in mostly repeated charts, such as bar charts, pie charts, and line charts, and very few maps. We checked the Financial Times Visual Vocabulary\footnote{https://www.ft.com/content/c7bb24c9-964d-479f-ba24-03a2b2df6e85} and Tableau Visual Vocabulary\footnote{https://www.tableau.com/solutions/gallery/visual-vocabulary}. These were useful to ensure we had a consistent vocabulary, but also showed several, very specific charts such as arc charts (typically shown only for election results), or waterfall charts.
We also visited certain specialized websites by conducting a search for "most common charts" (e.g., the now Atlassian-owned chart.io website \footnote{https://chartio.com/learn/charts/essential-chart-types-for-data-visualization/}, or Piktochart\footnote{https://piktochart.com/blog/types-of-graphs/}) and the results were quite limited, since they contained not-so common charts, such as funnel charts, or very specific charts such as Gantt charts, but were lacking in charts such as choropleths, very common in news articles. 
We also analyzed webpages that depicted relevant charts to analyze from New York Times from 2020\footnote{https://www.nytimes.com/2020/06/10/learning/over-60-new-york-times-graphs-for-students-to-analyze.html} and 2023\footnote{https://www.nytimes.com/2023/07/26/learning/over-75-new-york-times-graphs-for-students-to-analyze.html}. 

Finally, we evaluated the available charts in a popular tool, Datawrapper. The set of charts that are available in that location bears resemblance to the previous sets, shares the same vocabulary, and is sufficiently extensive. We discarded multiple charts' configurations and very specific charts, such as the arrow plot or the election donut, and added some widely known charts, such as violin plots.
As a result, we ended up with the following list of charts (the names are in alphabetical order, and they also represent the text used to prompt the LLM):

\begin{itemize}
 \item  Area chart: It uses a temporal, a categorical, and a quantitative dimension. 
 \item  Bar chart: Common bar charts, although they are assumed to be displayed horizontally in some sites (to distinguish them from vertical bars, that are called column charts). The data consists of a categorical variable and a quantitative one.
 \item  Box plot: The well-known chart to analyze data distributions. It requires a categorical variable and a quantitative one.
 \item  Bubble chart: Classical scatterplot with two quantitative dimensions, and a third one modulating the size of the dots.
 \item  Bullet chart: A bar chart with one or more reference values, typically depicted as a thinner and more opaque bar to indicate the value and a background bar to indicate the reference. It uses a categorical variable, plus two quantitative ones. 
 \item  Choropleth: Common map with colors encoding quantitative values for regions. It requires a geojson/json file with the geometry, and another file that contains a categorical data indicating the regions, and a quantitative value. 
 \item  Column chart: Typical bar chart. This name is used in contrast to bar to indicate that bars are vertical. Like in the previous case, both a categorical and a quantitative variable are required.
 \item  Donut chart: Similar to a pie chart. It represents data with one category and one quantitative dimension.
 \item  Dot plot: Scatterplots, though the name is often used to indicate that the variable encoded in Y may be discrete. It requires two quantitative values. 
 \item  Graduated symbol map: Map that encodes a quantity with a symbol (commonly a circle) that changes its size according to a quantitative variable. Like a choropleth, it uses a geojson/json indicating the regions and another CSV file with a categorical variable and one quantitative value.
 \item  Grouped bar chart: Same as column chart, but assumed that bars are horizontal. In this case, two categories and one quantity are represented. 
 \item  Grouped column chart: Same as bar chart, but emphasizing the vertical direction of the bars. Two categories and one quantity.
 \item  Line chart: With temporal, category, and quantitative variables.  
 \item  Locator map: Map where data encodes only a position given in latitude and magnitude. The symbol to denote the position can be a circle or a pin. Together with a geojson/json representing the regions, another map indicating the locations in GPS coordinates with a categorical variable (name of the object), together with longitude and latitude are required.
 \item  Pictogram chart: Also known as an isotype chart. It is similar to a quantized bar chart, where the length of the bar is formed by a series of isotype elements. It is encoded through a categorical and a quantitative dimension.
 \item  Pie chart: One category and one value are enough.
 \item  Pyramid chart: Used commonly for population data. Three variables are required, one categorical that indicates the ranges, and two quantitative variables for both the components.
 \item  Radar chart: It can be encoded with an ordered variable, a categorical one, and a quantitative one.  
 \item  Range plot: A categorical and a quantitative dimension are used. In this case, more than one quantity per category is required. 
 \item  Scatterplot: Two quantitative values.
 \item  Stacked bar chart: Like grouped bar charts, but different layout. Two categorical attributes and a quantitative one are necessary.
 \item  Stacked column chart: Identical to the previous one.
 \item  Violin plot: A more complex representation of distributions. It uses one category and one quantitative dimension.
 \item  XY Heatmap chart: Also known as heatmap matrix, or reorderable matrix, as named by Bertin \cite{bertin1973semiologie}. It uses two categorical (or ordered) dimensions in addition to a quantitative value that is encoded using a color.
\end{itemize}

\begin{table*}[ht]
  \caption{Analysis of charts that can be generated using ChatGPT3 and ChatGPT4. For this experiment, only the chart was prompted, and no library was imposed. Even though the same libraries are used most of the time, ChatGPT4 generates better outputs. It is noteworthy that ChatGPT4 utilizes seaborn in a lesser number of instances compared to ChatGPT3. }
  \label{tab:charts_validity}
  \scriptsize%
	\centering%
  \begin{tabu}{%
	r|cl|cl%
	}
  \toprule
   Technique & GPT3 & Selected library & GPT4 & Selected library \\
  \midrule
\textbf{Area chart} & \cmark~     & matplotlib and seaborn        & \cmark~ & matplotlib  \\
\textbf{Bar chart} & \cmark~     & matplotlib                    & \cmark~ & matplotlib \\
\textbf{Box plot} & \cmark~      & matplotlib and seaborn        & \cmark~ & matplotlib and seaborn \\
\textbf{Bubble chart}        & \cmark~ & matplotlib              & \cmark~ & matplotlib \\
\textbf{Bullet chart}       & \xmark~ & matplotlib               & \xmark~ & matplotlib \\
\textbf{Choropleth}          & \cmark~ & matplotlib              & \cmark~ & matplotlib \\
\textbf{Column chart}        & \cmark~ & matplotlib              & \cmark~ & matplotlib \\
\textbf{Donut chart}         & \cmark~ & matplotlib               & \cmark~ & matplotlib \\
\textbf{Dot plot}            & \xmark~ & matplotlib               & \cmark~ & matplotlib \\
\textbf{Graduated symbol map} & \xmark~ & matplotlib              & \xmark~ & matplotlib\\
\textbf{Grouped bar chart}   & \xmark~ & matplotlib               & \cmark~ & matplotlib \\
\textbf{Grouped column chart} & \xmark~ & matplotlib              & \cmark~ & matplotlib \\
\textbf{Line chart}          & \cmark~ & matplotlib and seaborn  & \cmark~ & matplotlib  \\
\textbf{Locator map}         & \cmark~ & matplotlib              & \cmark~ & matplotlib\\
\textbf{Pictogram chart}     & \xmark~ & matplotlib               & \xmark~ & matplotlib\\
\textbf{Pie chart}          & \cmark~ & matplotlib              & \cmark~ & matplotlib \\
\textbf{Pyramid chart}       & \cmark~ & matplotlib              & \cmark~ & matplotlib \\
\textbf{Radar chart}         & \xmark~ & matplotlib               & \xmark~ & matplotlib \\
\textbf{Range plot}          & \xmark~ & matplotlib and seaborn   & \xmark~ & matplotlib \\
\textbf{Scatterplot}         & \cmark~ & matplotlib              & \cmark~ & matplotlib \\
\textbf{Stacked bar chart}   & \cmark~ & matplotlib              & \cmark~ & matplotlib \\
\textbf{Stacked column chart} & \cmark~ & matplotlib            & \cmark~ & matplotlib \\
\textbf{Violin plot}         & \cmark~ & matplotlib              & \cmark~ & matplotlib and seaborn\\
\textbf{XY Heatmap chart}    & \cmark~ & matplotlib and seaborn  & \cmark~ & matplotlib and seaborn \\
\midrule
\textbf{Total}              & 16 ($66.7\%$) &                           & 19 ($79.17\%$) & \\
  \bottomrule
  \end{tabu}%
\end{table*}

\begin{figure*}[h!]
    \centering
  \begin{tabular}{cc}
      \includegraphics[width=0.54\linewidth]{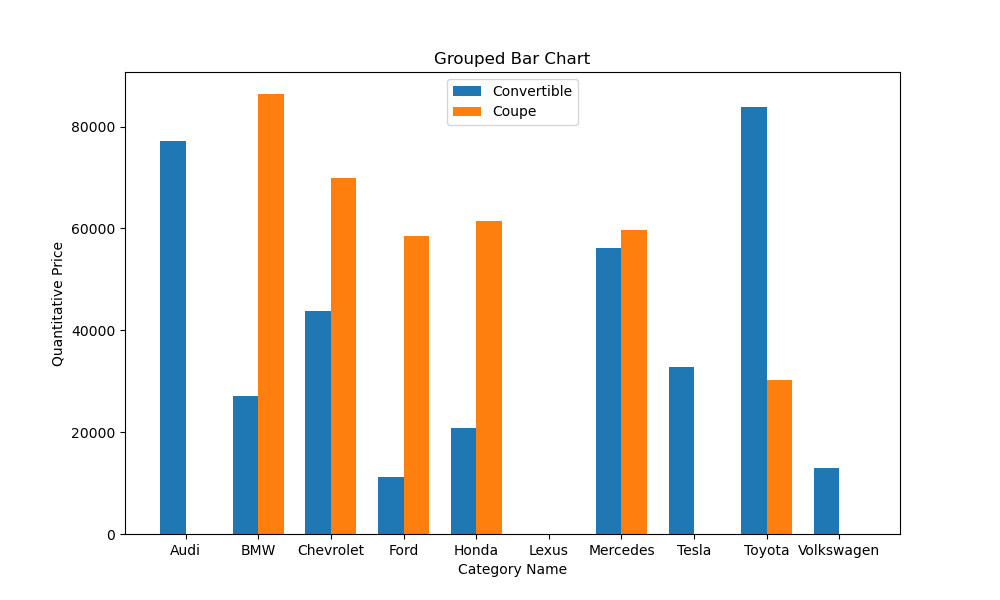} &  
      \includegraphics[width=0.45\linewidth]{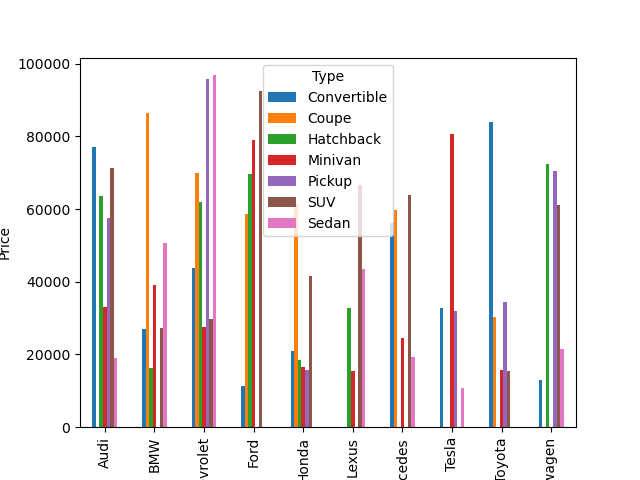} \\ 
       ChatGPT3 + matplotlib & ChatGPT4 + matplotlib \\ 
  \end{tabular}
    \caption{Generation of a grouped bar chart using ChatGPT3 and ChatGPT4 with the default configuration. In the first case, the system fails because it incorrectly assumes that only two values will be present per type category. ChatGPT, although it also uses matplotlib by default, properly generates as many bars as values per category. Nonetheless, the code additionally utilizes default values for the majority of the chart, resulting in unintended outcomes such as a legend that overlaps the data and labels that do not fit within the window.}
    \label{fig:GroupedBarCharts}
\end{figure*}

These charts include commonly known categories, except for hierarchical or network representations, which are difficult to see in most libraries. Moreover, those charts are very common in scientific literature, despite using sometimes slightly different names. 
As noted, there are some redundant charts, such as the bar vs column techniques. In data visualization, bar charts are assumed to be vertical (e.g., see \cite{munzner2014visualization}), although some Visual Vocabulary guides assume they are horizontal, or that bars refer to both (e.g., \cite{wilke2019fundamentals}). In general, unless otherwise specified, visualization libraries will generate vertical bars for both names. 

\subsection{Dataset creation}

The techniques depicted above have at least three different variables, which can be categorical, quantitative, temporal, or geometrical. In some cases, GPS location is needed, such as in the Locator maps. Hence, we needed specific files for testing the data.
We either borrowed or created several files. The following is a description of their contents:

\begin{itemize}
    \item myfile.csv: Contains five entries of a categorical variable and a quantitative variable. Suitable for simple charts.
    \item heatmapdataOrig.csv: Stores random values from 1 to 100 in a $50 \times 50$ matrix for the heatmap plots.
    \item iowa-electricity.csv: Sample file from the Vega Datasets, used for line data such as line charts or area charts. A shorter version has been derived for the radar chart.
    \item mycarsUnique.csv: A synthetic dataset with 53 entries that have two categories (trademark and type of vehicle) and a quantitative column indicating the price. Suitable for grouped and stacked bar charts.
    \item carsMod.csv: Based on a cars dataset, with multiple quantitative variables for bubble charts.
    \item myfileDotPlot.csv: A synthetic dataset with a categorical column and two quantitative values, for dot plots.
    \item population2021.csv: A derived dataset with the population of countries in 2021, with the name, abbreviation, year, and population. It has 274 entries, and it is used for the choropleth and graduated symbol map plots.
    \item pyramiddata.csv: Synthetic data with a first column indicating age ranges (17 different), and invented values of male and female populations in the next two columns.
\end{itemize}

In addition, we used two geojson files for the cartography of the world and the US. 

\subsection{Selection of LLMs}

Before starting the analysis, we performed some tests with different LLMs, including ChatGPT3 \cite{floridi2020gpt}, ChatGPT4 \cite{openai2023gpt4}, CodeLlama \cite{roziere2023code}, Mixtral, etc. We tested different versions of Llama2 and Code Llama2 (from Meta) and Mixtral (one of the open-source models from Mistral AI \footnote{https://mistral.ai/news/announcing-mistral-7b/}), were downloaded from Meta and HuggingFace, respectively. Locally, we tested the 7B parameter models by utilizing both the command line with proper libraries, LM Studio (a tool for easy deployment and testing of LLMs), and Chat2Vis (chat2vis.streamlit.io). Based on our initial experiments, neither Llama2, Code Llama2, nor Mixtral were exhibiting satisfactory performance. On numerous occasions, the generated code displayed errors. It is likely that larger models would perform better, but a high-end machine is required. Instead, ChatGPT3 (running from openai.com) and ChatGPT4 (through bing.com/chat) performed significantly well. As a result, we focused on those. 

\subsection{Prompts design}

To design the prompts, we defined the following requirements: 
\begin{itemize}
    \item Prompts should be simple to reproduce. They should all have a uniform structure: requesting the creation of a certain chart, employing the columns indicated from a CSV file.
    \item Vocabulary should be similar to the one used by commercial tools (e.g., Tableau), since many practitioners have their own public blogs where they discuss visualization. And that data is likely to be used to train LLM models.
    \item We would not use the concrete names of columns in the files, but just simple names such as A, B, C and data types would be described (e.g., "A (categorical), and B (quantitative)"). 
\end{itemize}

\begin{figure}[h]
    \centering
      \includegraphics[width=\linewidth]{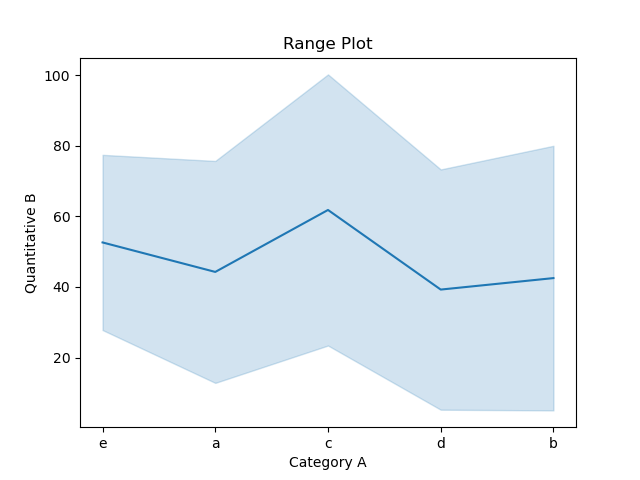}  
    \caption{Incorrect generation of a range plot by ChatGPT3. In this case, the chart is displayed, but it does not correspond to a range plot.}
    \label{fig:RangePlotChatGPT3}
\end{figure}

\begin{figure}[h]
    \centering
      \includegraphics[width=\linewidth]{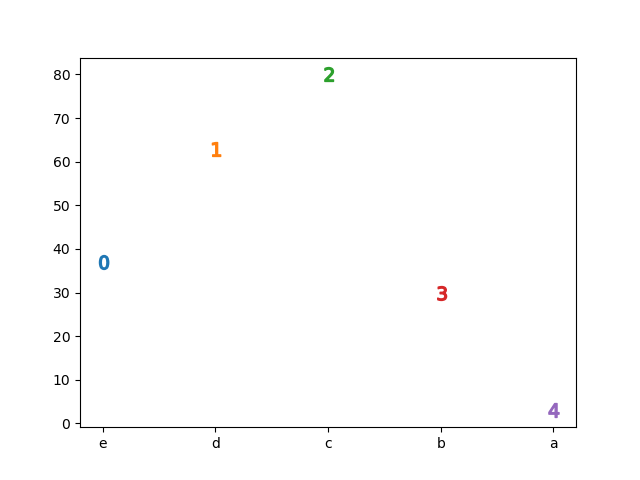}  
    \caption{Incorrect generation of a dot plot by ChatGPT4. The code executes, but the output resembles a scatterplot with numbers.}
    \label{fig:DotPlotChatGPT4}
\end{figure}

\begin{table*}[h]
  \caption{Results of the analyzed charts in ChatGPT4 using different libraries. In order to facilitate comparison, the results of the default configuration are provided as the second column. According to the experiment, no significant differences were found between libraries. The number of visualization techniques that are supported remains constant, although some of the techniques that are accurately represented may differ.}
  \label{tab:LibraryComparison}
  \scriptsize%
	\centering%
  \begin{tabu}{%
	l@{~~}l@{~~}l@{~~}l%
	}
  \toprule
   Technique & Matplotlib & Plotly (observations) & Altair (observations) \\
            & [+seaborn] &  &  \\
  \midrule
\textbf{Area chart}          & \cmark~ & \cmark~ & \cmark~\\
\textbf{Bar chart}           & \cmark~(categories not sorted) & \cmark~ (categories not sorted) & \cmark~ \\
\textbf{Box plot}            & \cmark~ (categories not sorted) & \cmark~  (categories not sorted) & \cmark~\\
\textbf{Bubble chart}        & \cmark~  (no legend for size) & \cmark~ (no legend for size, but interactive) & \cmark~\\
\textbf{Bullet chart}        & \xmark~ & \xmark~ & \cmark~ (with ticks)\\
\textbf{Choropleth}          & \cmark~ & \cmark~ (but legend not properly scaled) & \cmark~\\
\textbf{Column chart}        & \cmark~ (categories not sorted) & \cmark~ (categories not sorted) & \cmark~ \\
\textbf{Donut chart}         & \cmark~ (categories not sorted) & \cmark~ (categories not sorted) & \cmark~ (quantities on hover) \\
\textbf{Dot plot}            & \cmark~ (Y axis not sorted) & \cmark~ (Y axis not sorted) & \cmark~\\
\textbf{Graduated symbol map} & \xmark~ (generates choropleth) & \cmark~ & \xmark~ (expects centroids in data) \\
\textbf{Grouped bar chart}   & \cmark~ (clipped labels, legend overlaps) & \cmark~ & \cmark~\\
\textbf{Grouped column chart} & \cmark~ (clipped labels, legend overlaps) & \cmark~ & \cmark~ (horizontal array of horizontal bar charts) \\
\textbf{Line chart}          & \cmark~ & \cmark~ & \cmark~\\
\textbf{Locator map}         & \cmark~ & \cmark~ & \cmark~ (map inverted in Y) \\
\textbf{Pictogram chart}     & \xmark~ (generates scatterplot) & \xmark~ (not supported) & \xmark~ (not supported) \\
\textbf{Pie chart}           & \cmark~ (labels not sorted) & \cmark~ (labels not sorted) & \cmark~ (labels sorted) \\
\textbf{Pyramid chart}      & \cmark~ (legend overlaps) & \xmark~ (two pairs of bars) & \xmark~ (two bar charts side to side)\\
\textbf{Radar chart}        & \xmark~ (can be fixed manually) & \xmark~ (expects angles) & \xmark~ (not supported) \\
\textbf{Range plot}         & \xmark~ (expects different data) & \xmark~ (generates connected scatterplot) & \xmark~ (expects different data) \\
\textbf{Scatterplot}        & \cmark~ & \cmark~ & \cmark~\\
\textbf{Stacked bar chart}  & \cmark~ (clipped labels, legend overlaps) & \cmark~ & \cmark~  \\
\textbf{Stacked column chart} & \cmark~ (clipped labels, legend overlaps) & \cmark~ & \cmark~ (generates horizontal stacked bars)\\
\textbf{Violin plot}        & \cmark~ (categories not sorted) & \cmark~ (categories not sorted, points outside) & \cmark~\\
\textbf{XY Heatmap chart}	& \cmark~ (Y axis inverted) & \cmark~ (Y axis inverted) & \cmark~ (Y axis inverted)\\
\midrule
\textbf{Total}              & 19 ($79.17\%$) & 19 ($79.17\%$) & 19 ($79.17\%$)\\
  \bottomrule
  \end{tabu}%
\end{table*}

\subsection{Testing procedure}

For each experiment, we performed the following way:

\begin{itemize}
    \item Create a fresh new session.
    \item Input all the prompts in the same session and the same day.
    \item Copy the output code and edit the names of files and columns accordingly.
    \item Execute the Python code and analyze the results.
\end{itemize}

Since LLM chat sessions use the previous prompts as input, it was important to ensure that the session was created as a new one for each experiment. Hence, the vanilla tests, which do not require a specific library, and each library test are executed in distinct sessions. If many queries are generated asking for a chart in Altair and then perform a query without a library requirement, the output will be in Altair, instead of using matplotlib, which is the default library both ChatGPT3 and ChatGPT4 choose when asked for a chart in Python.

Furthermore, we wanted to keep the prompts as similar as possible. That is why we did not use real column names from the files. This ensures that no previous knowledge is used by the LLM. In addition, we specify the types of columns in every prompt to ensure that every query is self-contained.

\section{Generation of charts using LLMs}
\label{sec:Generation of charts using LLMs}

\subsection{Default configuration}

The first test we carried out was to evaluate how both ChatGPT3 and ChatGPT4 performed in free form, without the specification of the library. The outcomes were quite promising. They are shown in Table~\ref{tab:charts_validity}, together with the libraries the LLM system has selected for the output. The symbol \cmark~ indicates that the chart was generated correctly, and the symbol \xmark~ indicates that the chart is either depicting a different technique, or the code has some errors. A more thorough analysis of the issues that may appear even in correct charts is performed in \ref{subsec:Library tests}.

ChatGPT3 was able to generate code for 16 out of the 24 charts, exceeding the $66\%$ of the charts. Some of the problematic cases are relatively surprising. The failure cases for ChatGPT3 are: \textit{a)} bullet charts overlay a point and a bar chart, \textit{b)} for dot plots, it gerenates a connected scatterplot, \textit{c)} graduated symbol maps are drawn as choropleths, \textit{d)} grouped bar charts (both types) do not show all the elements in one category, as shown in Figure~\ref{fig:GroupedBarCharts} (but ChatGPT4 does, using the same library), \textit{e)} the pictogram chart prompt generates a line of circles of different sizes,   \textit{f)} radar charts generate wrong code, and \textit{g)} range plots generate a chart that resembles a line chart with a confidence interval (see Figure~\ref{fig:RangePlotChatGPT3}). 
ChatGPT4 works properly, although both use the same library. 

ChatGPT4, on the other hand, exhibits excellent performance in 19 out of the 24 distinct charts. The majority of issues arise from charts that necessitate combining with other charts, such as the range plot, which can be constructed by overlaying a line chart with a scatterplot or a line chart with circular marks at the data points. More concretely, the failure cases are: \textit{a)} bullet chart, where the output is a small multiples of pairs of horizontal bars one on top of the other, without width or opacity changes, \textit{b)} the graduated symbol map generates a choropleth, \textit{c)} the pictogram chart, that outputs a dot plot with labels (as shown in Figure~\ref{fig:DotPlotChatGPT4}), \textit{d)} the radar and range plot charts exhibit errors in the code. In the first case, the issue can be resolved manually by modifying two references to columns. Nonetheless, in the second case, the data is expected in a more distinct format.

\begin{figure*}[h]
    \centering
  \begin{tabular}{ccc}
      \includegraphics[width=0.2\linewidth]{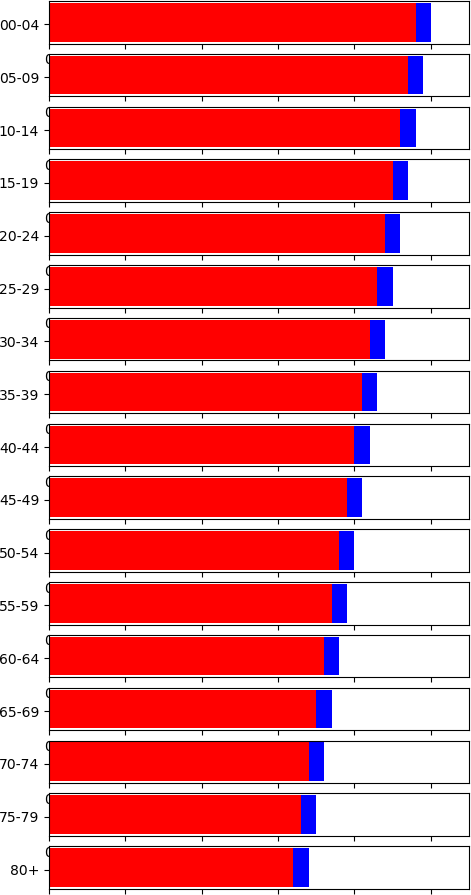} &  
      \includegraphics[width=0.42\linewidth]{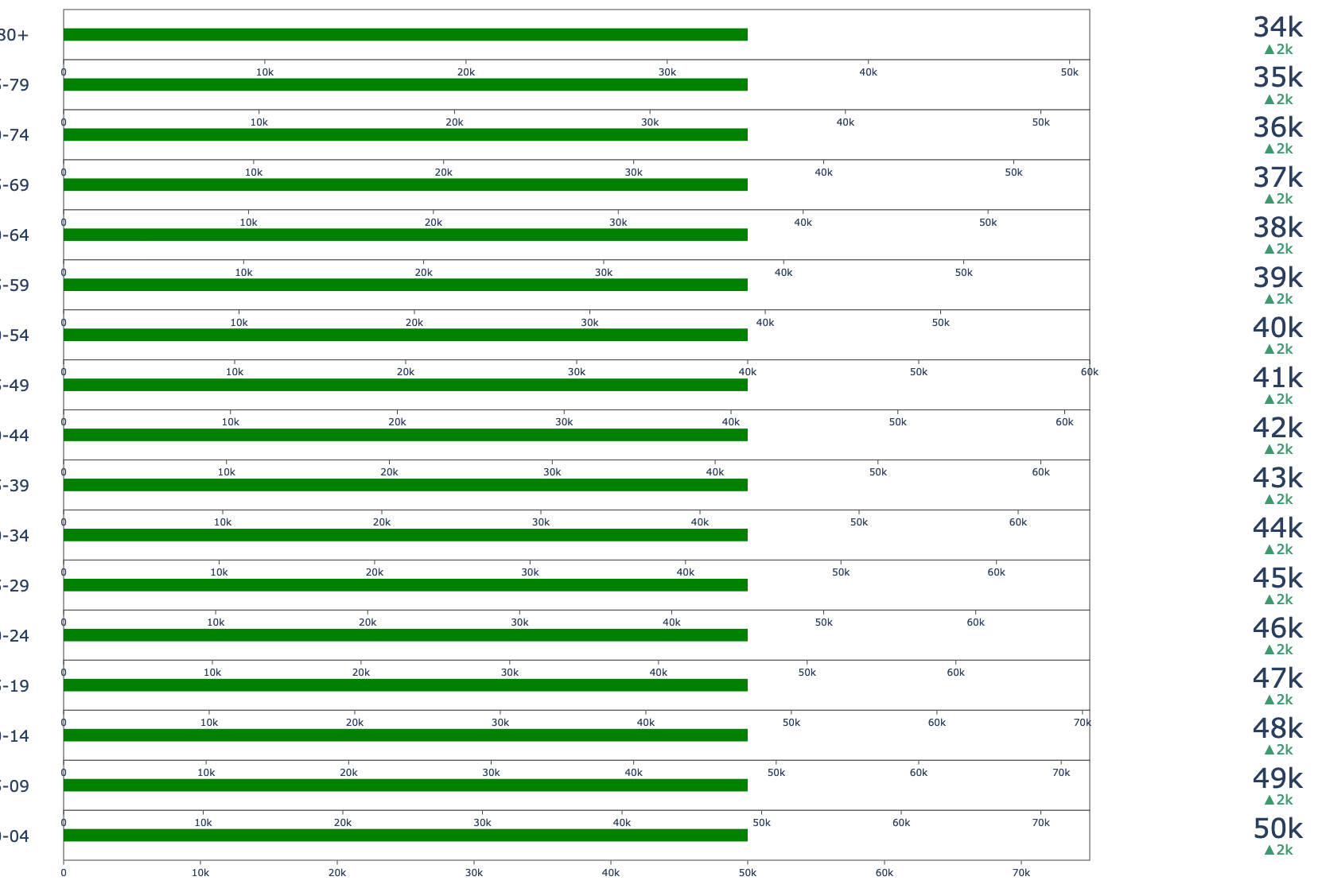} &
      \includegraphics[width=0.32\linewidth]{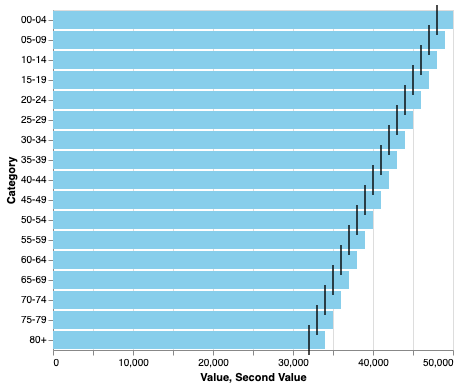} \\ 
       ChatGPT4 + matplotlib & ChatGPT4 + Plotly & ChatGPT4 + Altair \\ 
  \end{tabular}
    \caption{Generation of a bullet chart using the ages' dataset. The prompt specifies the column B as the value, and the column C as the second value. From the three prompts, only the one using Altair seems to succeed. The default configuration (left) renders two bars, but with the same size and opacity, thus making it difficult to understand whether it is a stack bar or two similar values are represented. Plotly generates something not very similar to a bullet chart, while Altair (right) uses a tick to mark the reference value.}
    \label{fig:BulletCharts}
\end{figure*}

\begin{figure*}[h]
    \centering
  \begin{tabular}{ccc}
      \includegraphics[width=0.28\linewidth]{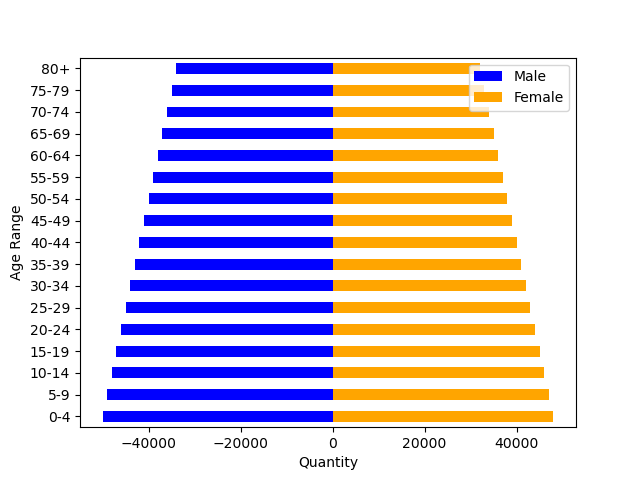} &  
      \includegraphics[width=0.29\linewidth]{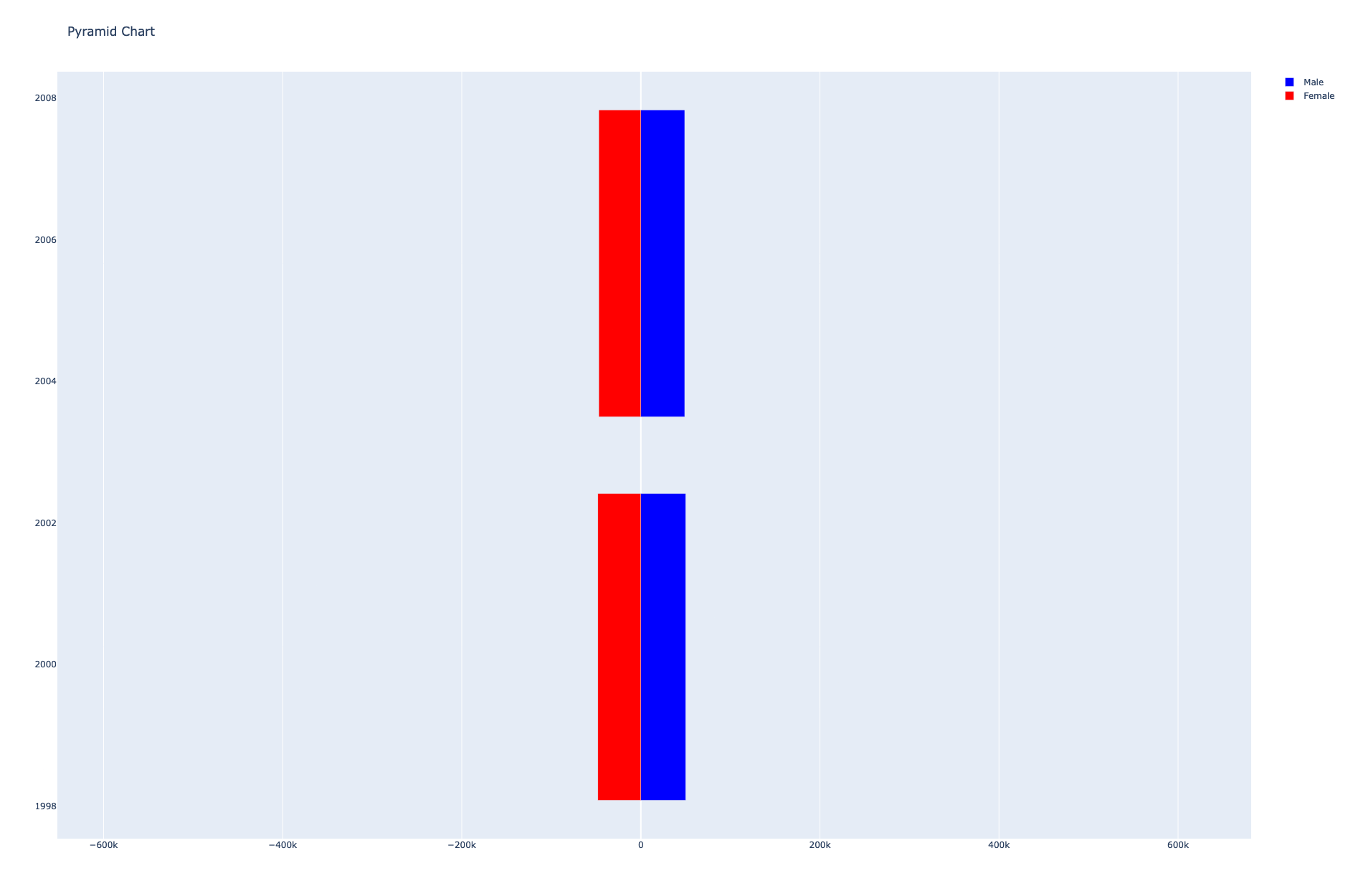} &
      \includegraphics[width=0.42\linewidth]{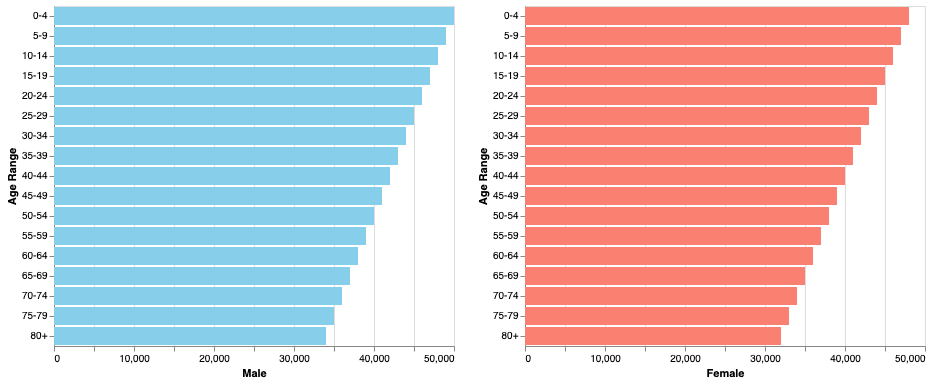} \\ 
       ChatGPT4 + matplotlib & ChatGPT4 + Plotly & ChatGPT4 + Altair \\ 
  \end{tabular}
    \caption{Generation of a pyramid chart using the ages' dataset. From the three prompts, only the one using the default ChatGPT4 configuration (left) works properly. Plotly generates something really awkward, while Altair (right) displays to side-by-side bar charts. }
    \label{fig:PyramidCharts}
\end{figure*}
\subsection{Library tests}
\label{subsec:Library tests}

The initial test indicated that ChatGPT generates charts, preferably using matplotlib. Nevertheless, there are many other popular Python libraries that can be used, such as bokeh, Plotlib, ggplot, Altair, and echarts. For the next test, we chose a well established library, Plotly \cite{sievert2020interactive}, and a relatively new one, Altair \cite{vanderplas2018altair}. Altair is a Python library that generates Vega Lite code \cite{satyanarayan2016vega}. Both libraries offer functionalities for interactive chart exploration; however, we shall solely examine the capabilities of generating static charts. 

To generate charts using a specific library, we added "with Plotly" or "with Altair" to the prompt. In all cases, the generated Python code used the desired libraries correctly. When imposing a library through the prompt, the results obtained with ChatGPT4 are similar to the previous ones. The number of charts that are now made correctly is the same for both libraries, 19 out of 24. However, there are some changes to the charts that are generated correctly and others that are coded incorrectly. As this experiment is more closely aligned with the requirements of both researchers and practitioners, we conduct a thorough analysis of the outcomes. Table \ref{tab:LibraryComparison} shows the results of the analysis. A check \cmark~ symbol indicates that the chart is correctly generated, despite it may have some presentation issues such as labels not fitting the window, or the legend overlapping the data. Most of these issues arise from the automation part of the library. Consequently, they are not considered to be failure cases. However, we provide a detailed account of the pertinent aspects that were discovered. The cross \xmark~ symbol indicates that the LLM did not produce the desired chart successfully. Either it produced a different chart, or the code breaks. Once again, the respective observation columns provide descriptions of the issues discovered.

By analyzing Table \ref{tab:LibraryComparison}, we can see that the behavior of different libraries is quite similar. There are some cases where they do not yield the same results, as in the bullet charts, illustrated in Figure \ref{fig:BulletCharts}. In this case, Altair seems to be the only library that performs as expected, since matplotlib and Plotly generate some output that does not correspond to the required chart. Another complex case is the one of pyramid charts. We expected better results, since population charts are quite common. However, only matplotlib performs well, as shown in Figure \ref{fig:PyramidCharts}.

\begin{table*}[h]
  \caption{Visual variables configurations using ChatGPT4 and the default library. The experiments indicate that certain visual variables appear to be more straightforward to configure than others. It is somewhat surprising that numerous attempts to modify the default configuration of scatterplots and bubble charts were unsuccessful. The proposed changes are quite common. However, the LLM system might have issues when those changes involve numerous columns, as it happens in those cases. The other techniques performed very well, and the only change that did not work was the definition of the sector colors in a pie chart.}
  \label{tab:VisualVariables}
  \scriptsize%
	\centering%
  \begin{tabu}{%
	l@{~~}l@{~}c@{~}l%
	}
  \toprule
   Technique & Prompt & Result & Observations \\
  \midrule
\textbf{Bar chart}           & with horizontal bars & \cmark~ & \\
\textbf{Bar chart}           & with bars in light green & \cmark~ &  \\
\textbf{Bar chart}           & with bars of width 120 pixels  & \cmark~ &  \\
\textbf{Bar chart}           & with the bars of different colors & \cmark~ & Continuous palette \\
\textbf{Bar chart}           & with title "This is a bar chart generated by an LLM" & \cmark~ & \\
\textbf{Bar chart}           & with labels indicating the quantities on top of the bars   & \cmark~  &  \\
\textbf{Line chart}          & with all lines in purple & \cmark~ & \\
\textbf{Line chart}          & with lines of width 20 pixels  & \cmark~ & \\
\textbf{Line chart}          & with the opacity of the lines as 0.5 & \cmark~ & \\
\textbf{Line chart}          & marking the data points with circles  & \cmark~ & \\
\textbf{Line chart}          & with green lines and the data points encoded with purple squares & \cmark~ & \\
\textbf{Line chart}          & with dashed lines & \cmark~ & \\
\textbf{Pie chart}           & sorting the values from larger to smaller & \cmark~  &  \\
\textbf{Pie chart}           & sorting the values from smaller to larger & \cmark~  & \\
\textbf{Pie chart}           & with the labels in boldface  & \cmark~ &  \\
\textbf{Pie chart}           & with the title "Pie"  & \cmark~  &  \\
\textbf{Pie chart}           &  with a sequential color palette that depends on the B column  & \xmark~  & \\
\textbf{Pie chart}           & sorting the values from larger to smaller, clockwise, and starting at 90 degrees & \cmark~  &  \\
\textbf{Pie chart}           & sorting the values from larger to smaller, clockwise, and starting at 90 degrees, & \cmark~  &  \\
                            & and not displaying the categorical labels, only the quantities, and outside the pie & \cmark~  &  \\
\textbf{Bubble chart}        & with points as triangles, whose angle depends on column D & \xmark~ & Error \\
\textbf{Bubble chart}        & with the shape  of points depending on column B & \xmark~ & No points, not enough categories\\
\textbf{Scatterplot}        & with the shape of points depending on column B and their size dependent  & \xmark~ & No points\\
                              & on column D with the opacity encoding the values in E column &  & \\
\textbf{Scatterplot}        & using columns C and D for X and Y axis and the size defined by the A column & \xmark~ & Big blue square\\
\textbf{Scatterplot}        & with the shape of points depending on column B and their size dependent on  & \xmark~ & No points\\
                                        & column D, and a size of 20 pixels &  & \\
\midrule
\textbf{Total}              & & 19 ($76\%$) & \\
  \bottomrule
  \end{tabu}%
\end{table*}

By examining the code, we saw that most of the chart configuration (legend position, sizes of fonts, etc.) were left to the library. As a result, on numerous occasions, this generates issues with the result, such as legends overlapping the data, or too tiny fonts. We succinctly described those issues in the respective observations columns. 

\subsection{Visual variables configuration}

To complete the experiments, we decided to include some prompts that, along with the chart creation, implied some modifications of the visual variables. For this experiment, we selected a subset of plots: bar charts, line charts, scatterplots/bubble charts and pie charts. To generate the chart, we modified the prompt with instructions to change some default parameters. We tested both changing visual variables based on fixed values (e.g., a certain width or opacity) and making some variables dependent on other columns. These changes refer to visual aspects of the marks, including direction, size, width, color, shape, opacity, stroke, ordering of the marks. In addition, we also introduced changes to the annotations: extra labels, their position, the font (boldface), and title of the chart. 

For each chart type, a different set of configurations was tested, based on common modifications we all made to our depictions. In Table \ref{tab:VisualVariables}, we show the modification prompt that was included in the chart creation, as well as the results. As it can be seen, most of the changes were successful: the code generated without errors, and the results were satisfactory. A notable exception was in the case of scatterplots and their variations. Most of the configurations generated either incorrect code or blank plots. We hypothesize that this might be due to the large number of variables included in the prompt. Since these plots were the ones that required a large number of visual variables' configuration changes dependent on input data.

\begin{figure*}[ht]
    \centering
  \begin{tabular}{ccc}
      \includegraphics[width=0.33\linewidth]{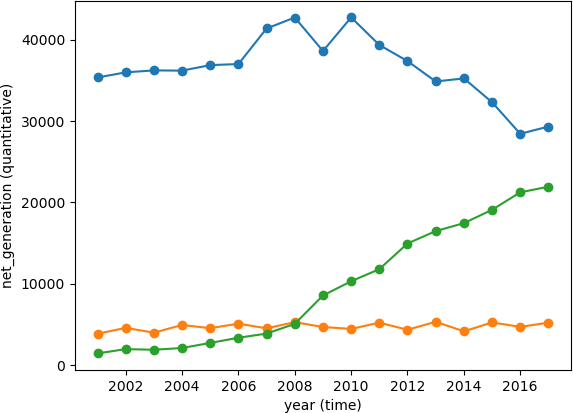} &  
      \includegraphics[width=0.33\linewidth]{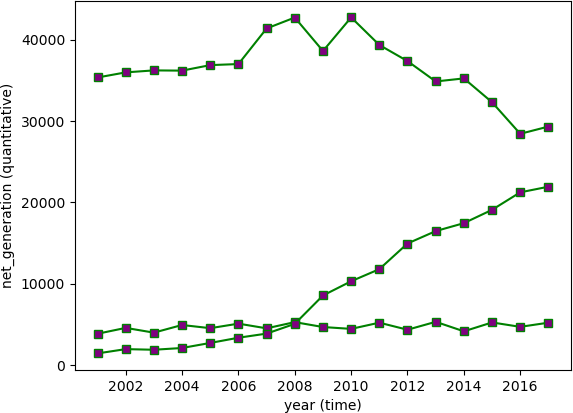} &
      \includegraphics[width=0.32\linewidth]{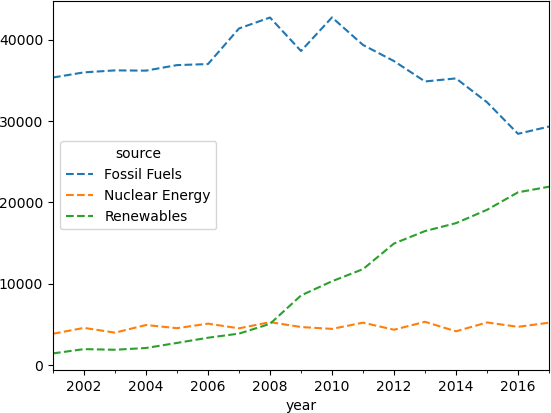} \\ 
       Adding points & Adding points as colored squares & Dashed lines \\ 
  \end{tabular}
    \caption{Configuring visual variables in line charts.}
    \label{fig:VisualVariablesLines}
\end{figure*}

\begin{figure*}[ht]
    \centering
  \begin{tabular}{cc}
      \includegraphics[width=0.4\linewidth]{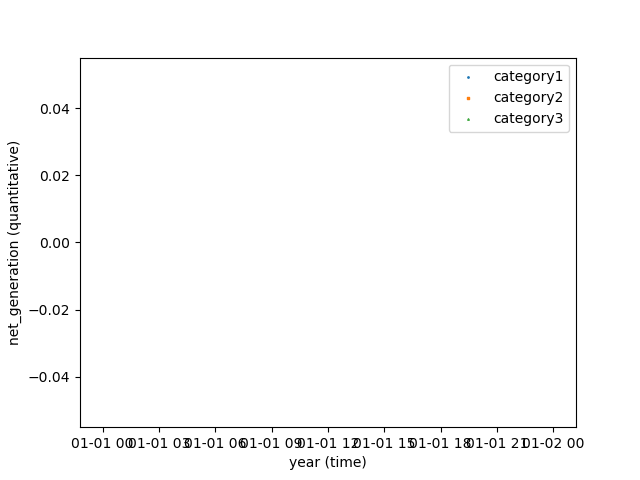} &  
      \includegraphics[width=0.4\linewidth]{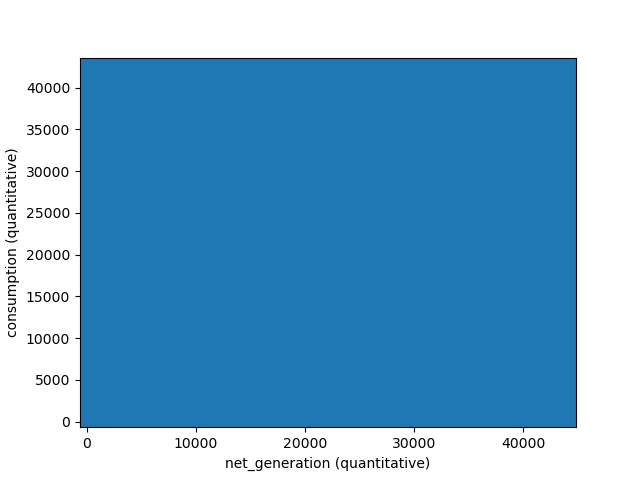}  \\ 
       Shape depending on column B & Using columns C and D for X and Y axes and A for the size 
  \end{tabular}
    \caption{Failure cases for scatterplot (left) with shape defined as a variable and bubble chart (right) with multiple visual variables defined.}
    \label{fig:VisualVariablesFail}
\end{figure*}

Some examples of successful prompts are shown in Figure \ref{fig:VisualVariablesLines} for the lines chart. Those prompts indicated changes in the marks, adding points as circles (left), adding points as squares with a different color (center), and changing the lines to dashes (right). As said, scatterplots and bubble charts are likely to fail when configuring multiple variables, as shown in Figure~\ref{fig:VisualVariablesFail}.

\section{Analysis of results}
\label{sec:Analysis of results}

The obtained results were very positive, and actually exceeded the author's expectations. We purposefully crafted single-prompt experiments, devoid of any subsequent inquiries. Even in the visual variables' configuration tests, the prompt was simple. This strategy is based on our belief that future LLMs will work well with a limited amount of input information. Furthermore, we believe that most users will expect a proper solution the first time. Hence, we sought to comprehend the extent of the capabilities of LLMs in those circumstances. Furthermore, we understand that follow-up prompts may improve the quality of the charts, as several papers have demonstrated. But we do not believe that the future for the users is to become prompt engineers. We are confident that advancements in LLM technologies will facilitate a more natural description of the output we aim to obtain from them. 

We also conducted further testing to see how the stochastic nature of LLMs may affect the results. For all the charts, we ran a test involving five distinct executions of the prompt using ChatGPT4, utilizing newly generated sessions for each prompt. In the subsequent executions, the charts that give errors are still not correct, with one exception: the bullet chart, which is generated correctly on one occasion. The rest of incorrect charts, were in all instances. For the charts that were initially correct, they are generated properly the rest of the executions, with the exception of the grouped bar chart, that is drawn wrong (inverted Y axis) in one out of the 5 attempts. Despite the charts getting the geometry correct, in three instances of different charts, a legend with a color palette that was unrelated to the representation was also incorporated to the code.

Besides, the experiments were performed on general LLMs. In the future, we expect that dedicated LLMs will also be broadly available, since fine-tuning for different tasks is a common step in neural model training. Thus, the results are only going to improve.  

The tests that we conducted were limited to make them manageable. There were various areas where we could have extended the resulting dataset. Initially, we restricted ourselves to predetermined prompts for each chart. Subsequently, the final prompts emerged from diverse experiments. Although some LLMs may have slightly better versions of such prompts, we do not expect any significant changes. We also limited ourselves in the number of techniques: as explained above, after analyzing both Financial Times and Tableau Visual Vocabulary, we found that some techniques are highly uncommon in newspapers and articles, and several require special data manipulation that is not simple for a non-specialist. Therefore, we finally took the simple charts available on Datawrapper.io and added some additional ones that were not there. Consequently, more precise techniques such as Sankey diagrams are absent, and even though they can be properly generated by LLMs, we believed that the sample was sufficiently representative. We also did not include interaction capabilities in the analysis. Again, this is something we have experimented with, using LLMs, but, describing interaction verbally is difficult, and believe that non-professionals might find it difficult to express naturally.

\section{Conclusions and Future Work}
\label{sec:Conclusions}

In this paper, we evaluated the capabilities of modern LLM systems to generate Python code that renders different visualization techniques. We have selected a total of 24 different techniques that are commonly used in research papers and infographics. We also designed a set of prompts to generate different techniques based on the vocabulary commonly used by professionals in the production or media industries. After conducting some tests, we decided to discard some publicly available 7B parameter models and opt for the public models of ChatGPT3 and ChatGPT4. We further analyzed the capabilities of generating Python code for different libraries by systematically evaluating these two different LLM models. We have decided to request the LLM systems to generate Python code instead of relying solely on visualization generation, as this approach enables us to gain a more profound understanding of the systems' output. Especially in the case of wrong code, we have had the possibility of correcting it or getting insights on the problems.

The results obtained were a positive surprise. Both the high number of charts that were properly built, and the fact that some of them are not straightforward, such as the maps, that need to combine spatial information with data. This makes us feel optimistic regarding the near future. However, there are still several shortcomings that cannot be solved easily by non-experts, such as the charts where labels are clipped, or the legends overlap the data. These require either prompt fine-tuning or code editing. Therefore, we believe that the present output may serve as a suitable starting point for users possessing programming or visualization expertise. However, end-to-end, from prompt to visualization flows, still do not achieve flawless results in numerous instances. 

In the future, we would like to include more complex visualizations in the test dataset, such as using combinations of charts, interactions, etc. Additionally, we would like to include more publicly available LLMs. Nonetheless, the tests conducted with 7B parameter models such as LlamaCode or Mixtral yielded outcomes that were not comparable (in fact, significantly inferior) to those obtained with ChatGPT3 or ChatGPT4. We also considered utilizing a JavaScript library, such as D3, but ultimately decided that the present selection was more suitable, given the substantial number of programmers who utilize Python and its relatively smooth learning curve in comparison to JavaScript.

\acknowledgments{
The author would like to thank the anonymous reviewers for their fruitful comments. Supported by Ministerio de Ciencia e Innovación/AEI (PID2021-122136OB-C21 by 10.13039/501100011033/FEDER, UE).}

\bibliographystyle{abbrv-doi}

\bibliography{template}

\end{document}